\documentclass[aps,prb,twocolumn,amsmath,superscriptaddress]{revtex4}
\usepackage{graphicx}
\usepackage{color}
\usepackage{epstopdf}
\usepackage{xcolor}

\begin{document}

\title{Influence of non-equilibrium phonons on the spin dynamics of a single Cr atom}

\author{V. Tiwari}
\affiliation{Institut N\'{e}el, CNRS, Univ. Grenoble Alpes and Grenoble INP, 38000 Grenoble, France}

\author{K. Makita}
\affiliation{Institute of Materials Science, University of Tsukuba, 1-1-1 Tennoudai, Tsukuba 305-8573, Japan}

\author{M. Arino}
\affiliation{Institute of Materials Science, University of Tsukuba, 1-1-1 Tennoudai, Tsukuba 305-8573, Japan}

\author{M. Morita}
\affiliation{Institute of Materials Science, University of Tsukuba, 1-1-1 Tennoudai, Tsukuba 305-8573, Japan}

\author{S. Kuroda}
\affiliation{Institute of Materials Science, University of Tsukuba, 1-1-1 Tennoudai, Tsukuba 305-8573, Japan}

\author{H. Boukari}
\affiliation{Institut N\'{e}el, CNRS, Univ. Grenoble Alpes and Grenoble INP, 38000 Grenoble, France}

\author{L. Besombes}\email{lucien.besombes@neel.cnrs.fr}
\affiliation{Institut N\'{e}el, CNRS, Univ. Grenoble Alpes and Grenoble INP, 38000 Grenoble, France}
\date{\today}

\begin{abstract}

We analyse the influence of optically generated non-equilibrium phonons on the spin relaxation and effective spin temperature of an individual Cr atom inserted in a quantum dot. Using a three pulses pump-probe technique, we show that the spin relaxation measured in resonant optical pumping experiments strongly depends on the optical excitation conditions. We observe for an isolated Cr in the dark a heating time shorter than a few hundreds $ns$ after an initial high power non-resonant excitation pulse. A cooling time larger than a few tens of $\mu s$, independent on the excitation, is obtained in the same experimental conditions. We show that a tunable spin-lattice coupling dependent on the density of non-equilibrium phonons can explain the observed dynamics. Low energy excitation conditions are found where the Cr spin states S$_z$=$\pm$1 can be efficiently populated by a non-resonant optical excitation, prepared and read-out by resonant optical pumping and conserved in the dark during a few $\mu$s.

\end{abstract}

\maketitle

\section{Introduction}

Electronic spins of individual defects in semiconductors are a promising platform for the development of quantum technologies such as quantum computing or quantum sensing. Integrating many localized spins into large quantum networks remains however challenging. Several proposal have recently suggested that spins interfacing could be achieved via mechanical degrees of freedom \cite{Schuetz2015,Lee2017,Lemonde2018} through the coupling to the same mode of a mechanical oscillator or to propagating surface acoustic waves \cite{Schuetz2015}. In this context, development of spin $qubits$ in semiconductors based on localized spins with large intrinsic spin to strain coupling are particularly attracting.

Magnetic atoms in semiconductors featuring both spin and orbital degrees of freedom can present a large spin to strain coupling and the spin of an individual magnetic atom can be initialized and probed using the optical properties of a quantum dot (QD) \cite{Koenrad2011,Besombes2004,Kudelski2007,Goryca2009,LeGall2011,Krebs2013,Smolenski2016}.
This has been in particular recently demonstrated for Chromium (Cr) in a II-VI semiconductor QD which carries both an electronic spin S=2 and an orbital momentum L=2. The characteristic photo-luminescence (PL) spectra of Cr-doped QDs resulting from the exchange interaction between confined carriers and the electronic spin S=2 of an individual Cr atom were identified in magneto-optic measurements \cite{Lafuente2016}. A spin to strain coupling more than two orders of magnitude larger than for elements without orbital momentum (NV centers in diamond \cite{Tessier2014}, Mn atoms in II-VI semiconductors \cite{Lafuente2015} ... ) was evidenced.

In analogy with the spin structure of NV centers in diamond, the spin states S$_z=\pm1$ of a Cr in a self-assembled QD form a $qubit$ ($\{+1;-1\}$ spin $qubit$) strongly coupled to in-plane strain \cite{Lee2017}. It was also demonstrated that resonant optical pumping can be used to empty either the S$_z=+1$ or S$_z=-1$ spin states \cite{Lafuente2017Cr,Lafuente2018}. The incorporation of Cr in II-VI semiconductors is therefore a promising route for the realization of hybrid spin-mechanical devices in which the motion of a mechanical oscillator would be coherently coupled to the spin state of a single atom and probed or coherently controlled through this interaction \cite{Rabl2010,Tessier2014,Ovar2014}. Conversely, a dynamical strain field could be used for a coherent control of the $\{+1;-1\}$ Cr spin $qubit$. 

However, spin-phonon coupling is also the ultimate spin relaxation process for an isolated spin in a solid state environment \cite{Abragam}. Spin-phonon coupling is expected to be efficient for magnetic elements carrying an orbital momentum where a modulation of the crystal field by the strain field of phonons combined with the spin-orbit coupling can induce spin transitions. A spin relaxation time in the $\mu$s range was recently observed for an isolated Cr spin but the mechanisms controlling the spin relaxation were not identified in these optical pumping experiments \cite{Lafuente2017Cr,Lafuente2018}.

We show in this article that the optically measured spin relaxation time for the states S$_z=\pm1$ of a Cr atom in a CdTe/ZnTe QD is strongly influenced by non-equilibrium phonons generated during the optical excitation. After a short presentation of the QDs samples and spin structure in Cr-doped QDs, we detail the resonant optical pumping experiments and their particular dependence on the optical excitation power. We first demonstrate that the $S_z=\pm1$ spins states are efficiently populated by the interaction with non-equilibrium phonons generated during a high energy optical excitation. This permits to initialize the system in the wanted $\{+1;-1\}$ spin subspace to be used as a $qubit$. We then show how a three pulses pump-probe experiment can be used to probe the influence of optically created non-equilibrium phonons on the Cr spin dynamics. We observe for an isolated Cr in the dark a power dependent heating time which can be in the order a few hundreds $ns$ after an initial high intensity non-resonant excitation pulse. A cooling time larger than ten $\mu$s independent on the initial excitation is obtained with the same experimental conditions. We finally discuss the mechanisms responsible for the Cr spin dynamics and shown that a tunable spin-lattice coupling strongly dependent on the density of non-equilibrium phonons explains the observed behaviour.

\section{Samples and experiments}

The studied sample consists of a layer of Cr-doped self-assembled CdTe QDs grown by molecular beam epitaxy on a p-doped ZnTe (001) substrate \cite{Wojnar2011,Lafuente2016}. A low density of Cr in the CdTe layer is chosen to avoid multiple Cr inside \cite{Lafuente2018} or in the vicinity \cite{Besombes2019} of the dots to interact with the carriers confined in a given QD.

The PL of individual QDs was studied by optical micro-spectroscopy at low temperature (T$_0$=5K). The PL was excited with a continuous wave (CW) dye laser tuned to an excited state of the QDs, dispersed and filtered by a 1 $m$ double spectrometer before being detected by a Si cooled multichannel charged coupled device (CCD) camera or, for time resolved experiments, a Si avalanche photodiode with a time resolution of about 350 ps. The laser power was stabilized by an electro-optic variable attenuator. A single-mode dye ring laser could also be tuned on resonance with the ground state of the exciton coupled with Cr to perform time resolved resonant optical pumping experiments. Trains of resonant and non-resonant light pulses with variable delays and wavelengths were generated from the CW dye lasers using acousto-optical modulators with a switching time of about 10 ns. In all these experiments, a high refractive index hemispherical solid immersion lens with a diameter of 1 mm was deposited on top of the sample to enhance the collection of the single-dot emission.

For a Cr inserted in a self-assembled QD, a magnetic anisotropy induced by the bi-axial strain dominates the spin fine structure of the atom. It results from an strain induced modification of the crystal field and the spin-orbit coupling \cite{Vallin1974}. This magnetic anisotropy can be described by an effective spin Hamiltonian D$_0$S$_z^2$ where $z$ is the QD growth axis. With D$_0$ in the 1-2 meV range in a CdTe/ZnTe QD, this parabolic splitting shifts the spin states S$_z$=$\pm$2, and in a lesser extend the spin states S$_z$=$\pm$1, to high energy where they cannot be significantly populated at a lattice temperature T$_0$=5K. The Cr spin thermalizes on the lowest energy states and no contribution of the S$_z$=$\pm$2 spin states have been observed until now in the PL of Cr-doped dots in the investigated temperature range (from 5K to 40K). However, as it will be demonstrated in section III, the optical excitation significantly populates the spin states S$_z$=$\pm$1 through the generation of non-equilibrium phonons and spin-lattice coupling.

\begin{figure}[hbt]
\centering
\includegraphics[width=3.25in]{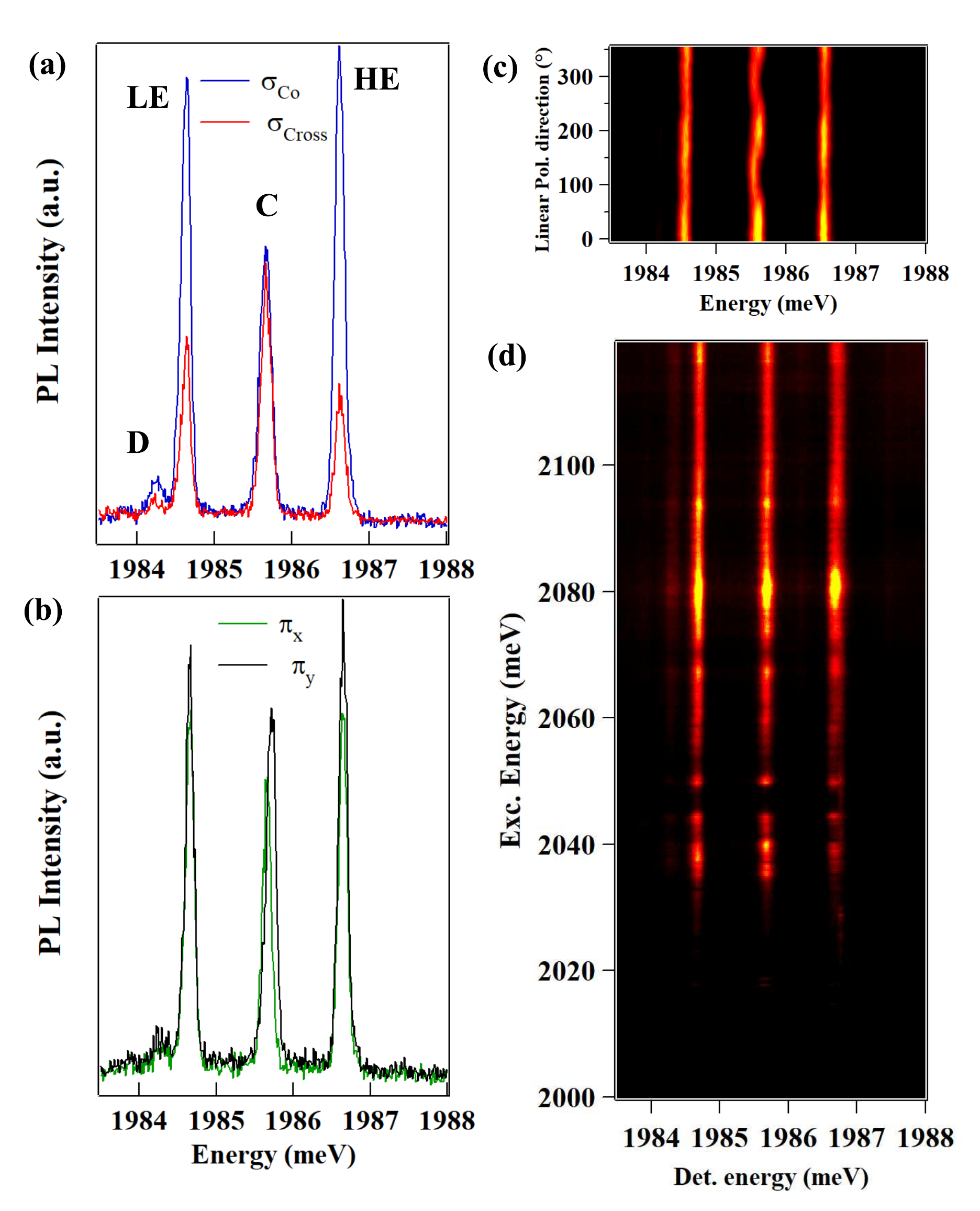}
\caption{PL characteristic of a singly Cr-doped CdTe/ZnTe QD measured at T=5K. (a) Co and cross circularly polarized PL spectra. (b) PL spectra recorded along two orthogonal directions x and y. (c) Linear polarisation intensity map. (d) Intensity map of the Photo-Luminescence Excitation spectra.}
\label{FigECr}
\end{figure}

The PL spectra of an exciton in an individual Cr-doped QD (X-Cr) is reported in figure \ref{FigECr}. It is dominated by three main emission lines. Dots containing a single Cr atom can be clearly identified by using circularly polarized excitation / detection and linear polarization analysis. For a circularly polarized excitation, the emission of the two outer lines (lines HE and LE) is co-polarized whereas the central line does not present any circular polarization. A linear polarization analysis (figure \ref{FigECr} (b) and (c)) shows that the central line is indeed a doublet linearly polarized along two orthogonal directions.

This PL structure is a consequence of the exciton-Cr exchange interaction. The latter splits the Cr spins states depending on their relative orientation with the spin of the heavy-hole bright excitons $|\Downarrow_h\uparrow_e\rangle$ or $|\Uparrow_h\downarrow_e\rangle$ \cite{Lafuente2016}. The two outer lines in the PL spectra (LE and HE in figure \ref{FigECr}) are then linked with the Cr spin states $S_z$=$\pm$1 and the central line (line C) corresponds to $S_z$=0 (see figure \ref{Fig2Sch}). The exchange interaction with the spin states S$_z=\pm$1 acts as a magnetic field on the exciton which can then conserve its spin during its lifetime and give co-polarized PL.

Most of the dots also present a low symmetry and the central line ($S_z$=0) is split by the electron-hole exchange interaction and linearly polarized along two orthogonal directions (see figure \ref{FigECr} (b) and (c)). This is the fine structure splitting usually observed in non-magnetic QDs. This splitting is responsible for the absence of circular polarisation on the central line. An additional line, labelled D in figure \ref{FigECr}, often appears on the low-energy side of the exciton PL spectra. It arises from a dark exciton ($|\Uparrow_h\uparrow_e\rangle$ or $|\Downarrow_h\downarrow_e\rangle$) which acquires some oscillator strength by a mixing with a bright exciton associated with the same Cr spin state \cite{Lafuente2016}. The fine structure splitting and the contribution of the dark exciton are very weak in the QD investigated here showing that it is close to a perfect cylindrical symmetry.

\section{Optical pumping and Cr spin effective temperature}

\begin{figure}[hbt]
\centering
\includegraphics[width=3.25in]{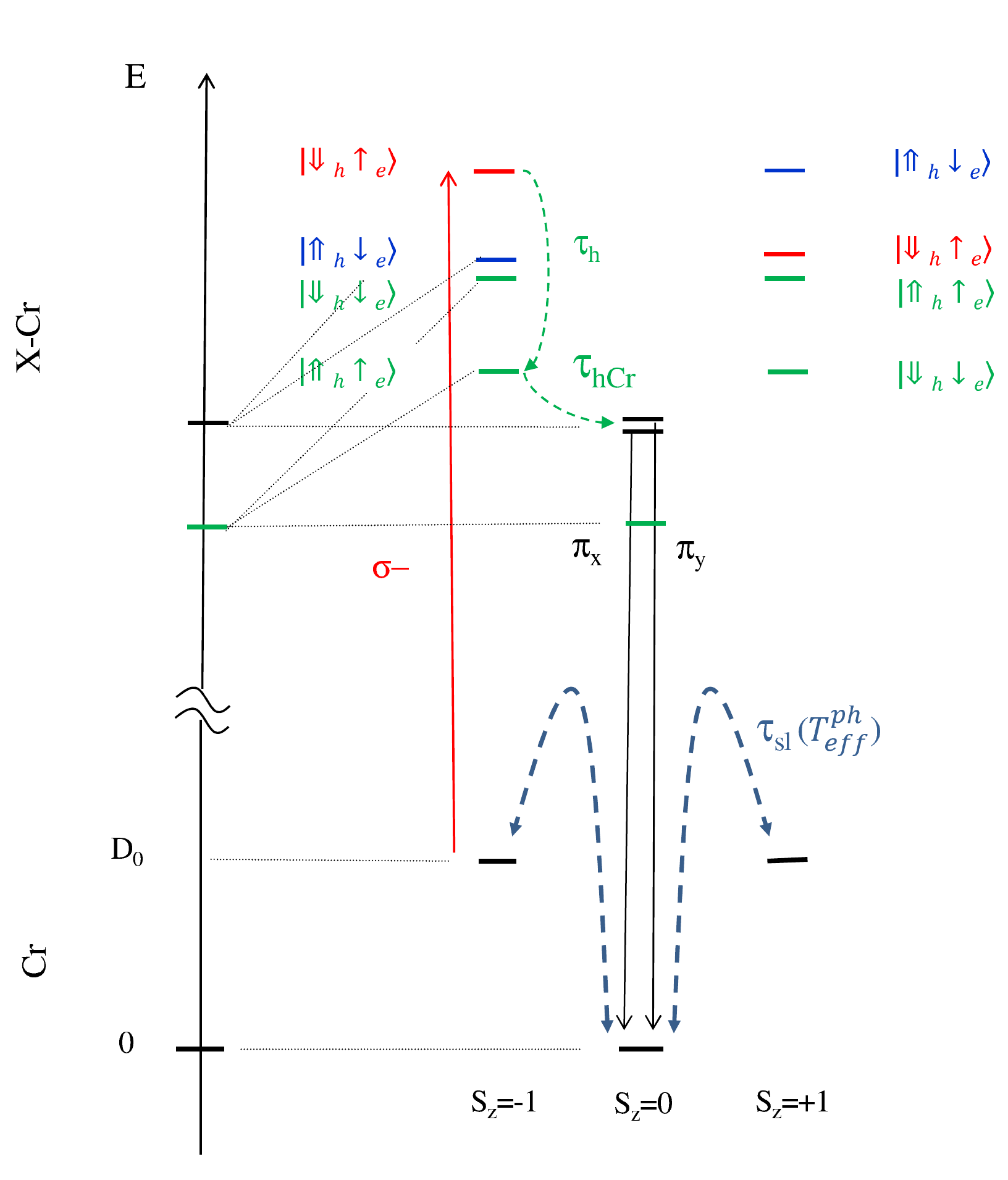}
\caption{Energy spin level structure of a Cr atom with (X-Cr) and without (X) the presence of an exciton in a CdTe/ZnTe QD. Only the populated states with S$_z$=0 or S$_z$=$\pm$1 are presented. $|\Uparrow_h\rangle$ and $|\Downarrow_h\rangle$ correspond to the heavy hole spin up and down while $|\uparrow_e\rangle$ and $|\downarrow_e\rangle$ illustrate the spin up and down of the electron. The optical and main spin relaxation paths involved in the optical pumping are presented for a $\sigma-$ resonant excitation on the HE line. A non-resonant excitation populates the spin state S$_z$=$\pm$1 via the generation of non-equilibrium acoustic phonons and the spin-lattice coupling $\tau_{sl}$(T$_{eff}^{ph}$). The hole-Cr flip-flop ($\tau_{hCr}$) transfers the Cr spin from S$_z$=-1 to S$_z$=0.}
\label{Fig2Sch}
\end{figure}

The spin states S$_z$=$\pm$1 of a Cr atom inserted in a QD can be pumped under resonant optical excitation \cite{Lafuente2017Cr}. This initialization of the Cr spin can be detected in the resonant PL of the QD. The main mechanism of optical pumping was recently identified \cite{Lafuente2018}: The spin flip of the Cr are induced by hole-Cr flip-flops resulting from an interplay of the hole-Cr exchange interaction and the interaction with the strain field of acoustic phonons during the lifetime of the excitons (see figure \ref{Fig2Sch}). These flip-flops can take place in a few $ns$ but their probability strongly depend on the energy difference between the exciton-Cr states involved in the transition.

Under resonant excitation on the high (or low) energy line of X-Cr with either S$_z$=+1 or S$_z$=-1, the flip-flop process induces a transfer of the exciton towards the central line associated with the spin state S$_z$=0 \cite{BesombesSST2019} (line C in figure \ref{FigECr}). In previous studies (reference \cite{Lafuente2018,BesombesSST2019}), dots with a weak exchange splitting were investigated and, because of the inhomogeneous broadening of the lines, a large intensity of the resonant laser was also used. The population transfer from S$_z$=$\pm$1 to S$_z$=0 was evidenced using cross-linear excitation / detection on lines HE and C respectively to limit the direct absorption in the acoustic phonon side-band of the central line \cite{Lafuente2018}.

Here, the selection of a Cr-doped QD with a large exchange induced splitting and a weak inhomogeneous broadening of the PL lines permits to perform resonant optical pumping experiments under cross-circular excitation and detection on the high energy line and central line respectively. The resonant PL transient recorded on C can then be used to efficiently probe the optical pumping of either the spin states S$_z$=+1 or S$_z$=-1 and analyse their dynamics.

\begin{figure}[hbt]
\centering
\includegraphics[width=3.5in]{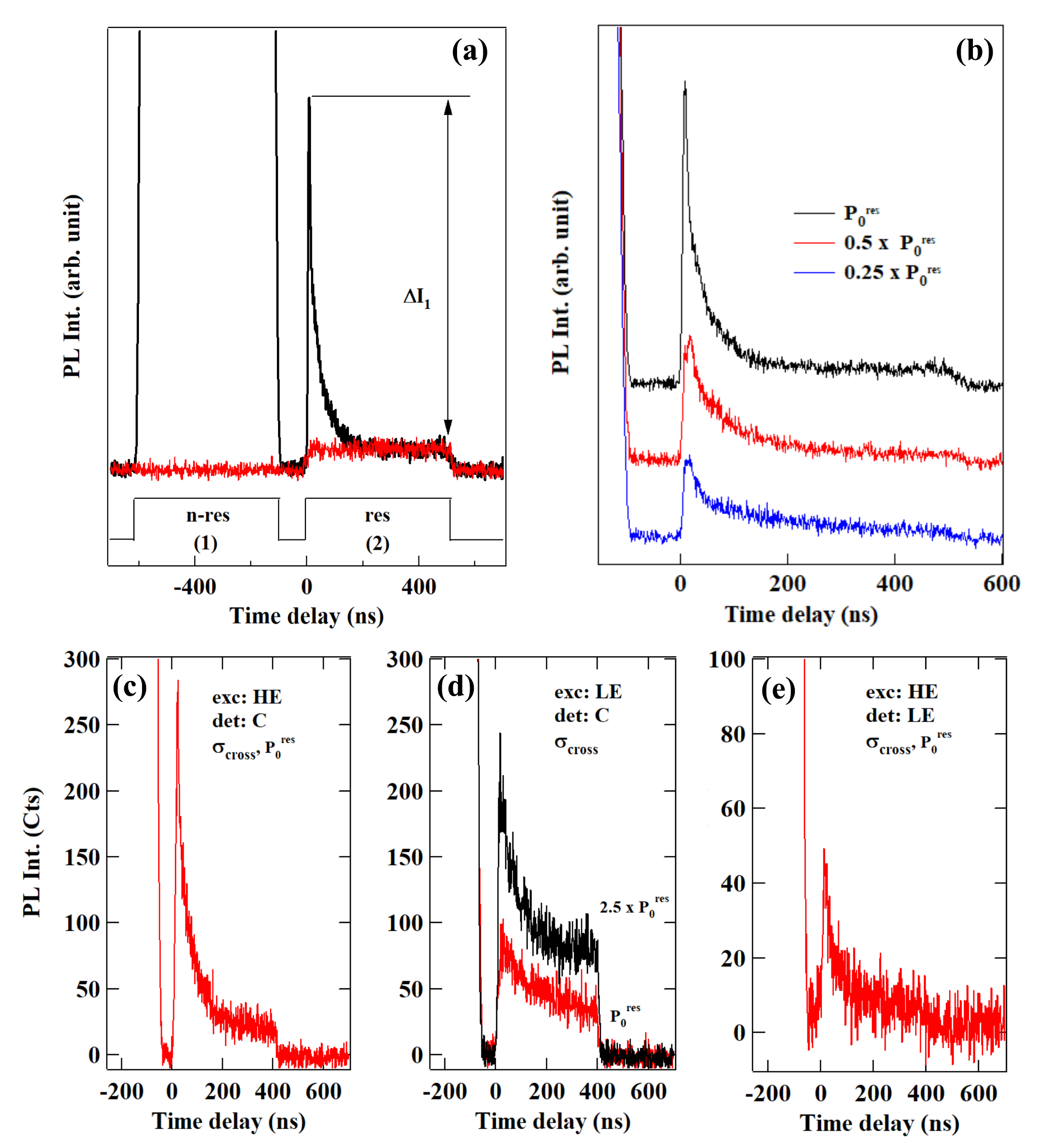}
\caption{(a) Resonant optical pumping transients recorded on the central line (C) for cross-circularly resonant excitation on the hight energy line (HE). (b) Resonant excitation power dependence (P$^{res}$) of the pumping transient. Comparison of the pumping transient recorded: (c) on the central line (C) for an excitation on the hight energy line (HE), (d) on the central line (C) for an excitation on the low energy line (LE) and (e) on the low energy line (LE) for an excitation on the high energy line (HE). Note the difference in the vertical scales of (e). The optical excitation sequence is shown in (a).}
\label{Fig2}
\end{figure}

The main features of the resonant optical pumping experiments are summarized in figure \ref{Fig2}. We use a two wavelengths experiment. A first pulse (pulse 1) with a 500 ns duration, tuned to high energy on an excited state of the QD (around 2080 meV, see figure \ref{FigECr}(d)) is used to increase the effective temperature of the Cr spin and significantly populate the spin states S$_z$=$\pm$1. A second circularly polarized resonant pulse (pulse 2), tuned to the high energy line of the QD is used to perform the optical pumping. A $\sigma$+ laser on the high energy line will address the spin state S$_z$=+1 and a $\sigma$- laser, S$_z$=-1. The dynamics of the resonant PL detected on C probes the speed of the initialization ({\it i.e.} emptying of the spins states +1 or -1) and the amplitude of the observed transient, $\Delta I_1$ in figure \ref{Fig2}, reflects the initial population of the spin states S$_z$=+1 or S$_z$=-1.

As depicted in figure \ref{Fig2}(a), the intensity of the resonant PL measured during pulse 2 reaches a weak intensity plateau after a large transient of amplitude $\Delta I_1$: the optical pumping is almost complete. The dynamics of the optical pumping depends on the intensity of the resonant laser and can take place in a few tens of $ns$ at high excitation power (figure \ref{Fig2}(b)). For a given resonant excitation power, the speed of the optical pumping is slower for an excitation on the low energy line than for an excitation on the high energy line (figure \ref{Fig2}(c) and (d)). This is a consequence of the mechanism of hole-Cr flip-flop whose efficiency depends on the energy splitting between the X-Cr levels involved in the spin transitions \cite{BesombesSST2019}.

Let us note that the resonant pumping signal could also be detected on the resonant PL of the low energy line for a cross-circularly polarized excitation and detection (excitation and detection of the same Cr spin state). However, this signal is coming from forbidden spin-flip of the exciton and is, for the QD studied here, significantly weaker than the one detected on C (figure \ref{Fig2} (e)). We will use for all the spin dynamics measurements presented in this paper a cross-circular excitation and detection on the high energy and central lines respectively.

\begin{figure}[hbt]
\centering
\includegraphics[width=3.0in]{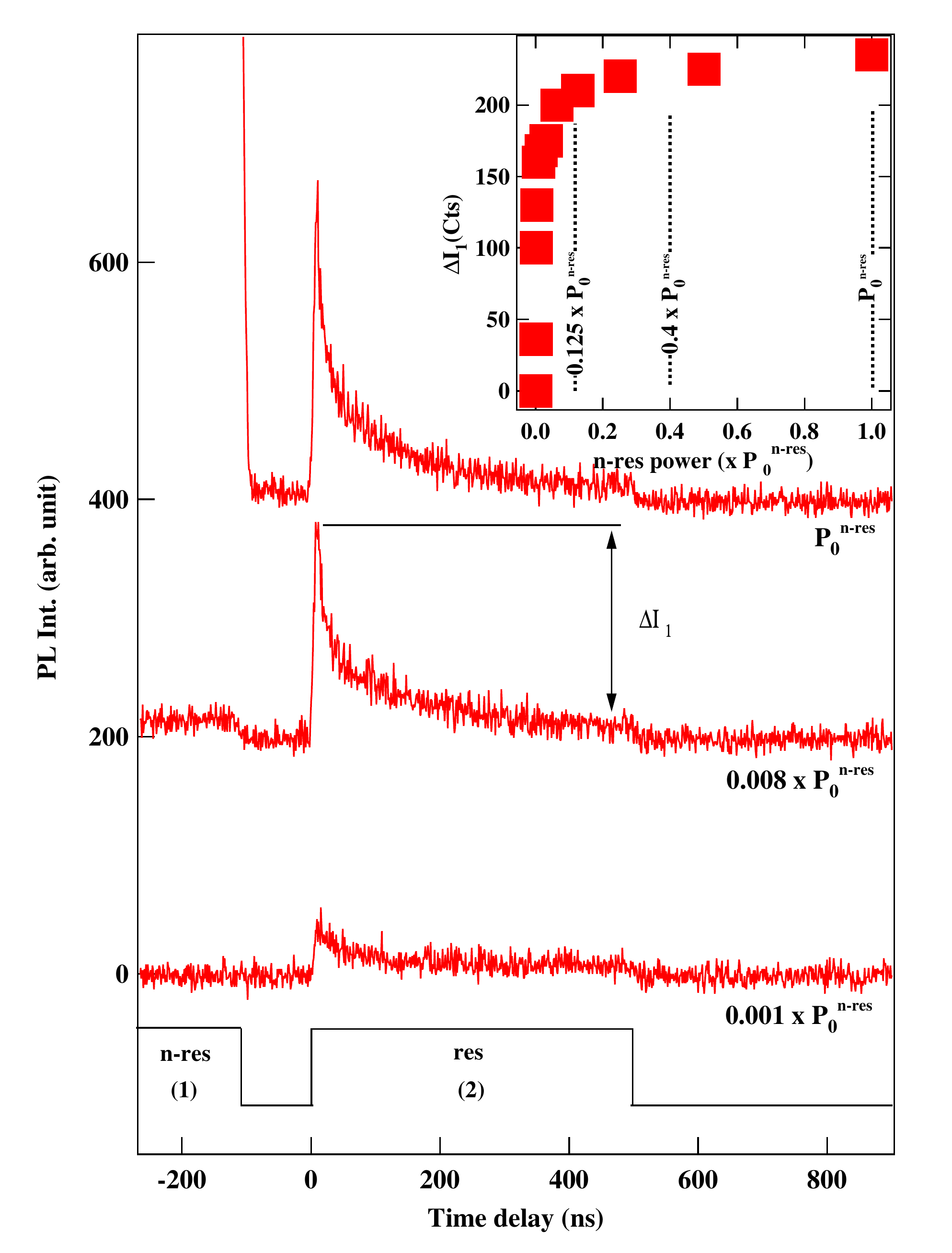}
\caption{Evolution of the resonant pumping signal as a function of the power of the non-resonant pulse 1, P$^{n-res}$. The inset shows the excitation power dependence of the amplitude of the transient $\Delta I_1$. Vertical lines indicate the excitation powers used in the time resolved optical pumping experiments reported in figure \ref{Fig4}.}
\label{Fig3}
\end{figure}

It was demonstrated (see reference \cite{Lafuente2018}), using a two beams experiment where the non-resonant pulse was tuned a few $\mu$m away from the studied QD, that the main contribution to the Cr spin heating under high energy excitation is due to the interaction with non-equilibrium phonons generated by the relaxation of photo-created high energy carriers \cite{Lafuente2018}. During the non-resonant pulse, the Cr spin reaches an effective temperature T$_{eff}^{Cr}$ much larger than the temperature of the lattice where the $\pm$1 spin states are significantly populated. This effective temperature depends on the power of the high energy excitation either if carriers are injected or not injected inside the QD \cite{Lafuente2018}.

Under high energy excitation on a semiconductor, the non-equilibrium phonon spectrum generated during the relaxation of carriers may be described as a distribution with energy dependent effective temperature $N_{\omega}=1/(e^{-\hbar\omega/k_BT_{eff}^{ph}(\omega)}-1)$.

In equilibrium, the phonon spectrum is Planckian and T$_{eff}^{ph}$ does not depend on $\omega$. In the excitation spot the phonon spectrum is non-equilibrium and T$_{eff}^{ph}(\omega)$ has a complex $\omega$ dependence. $N_{\omega}$ includes ballistic phonons from the laser excitation and equilibrium bath phonons with T$_0$=5K. T$_{eff}^{ph}(\omega)$ is difficult to estimate as it depends on the spectrum of phonons initially generated by the optical excitation and on the scattering of these ballistic phonons on defects or at the sample surface. Scattering can induce a decay of an initial high energy phonon into two phonons of lower energies. T$_{eff}^{ph}(\omega)$ is then expected to decrease at high $\omega$ and, for a fixed $\omega$, decrease with time after the end of the excitation pulse as phonons progressively thermalize or are evacuated from the sample at the interface with external medium.

The Cr spin temperature is mainly controlled by $N_{\omega}$ through the spin-lattice coupling. Here we chose however a direct excitation which also inject carriers inside the QD to use the large PL produced during the non-resonant excitation to track the position of the QD during the time resolved resonant pumping experiments.

The evolution of the effective temperature of the Cr can be observed in the power dependence of the amplitude of the resonant pumping signal performed just after the end of pulse 1 ($\Delta$I$_1$ in figure \ref{Fig3}). No detectable optical pumping transient is obtained if the non-resonant heating pulse 1 is suppressed. This shows that either there is almost no relaxation during the dark time between two resonant pulses (2 $\mu$s in this particular experiment) or that with this excitation condition the spin effective temperature of the Cr is too weak to obtain a significant population of the spin states $S_z$=$\pm$1 and observe a resonant PL transient.

With the increase of the intensity of the non-resonant excitation (pulse 1), the amplitude of the transient $\Delta$I$_1$ presents first a fast initial increase before it saturates rapidly at high power (inset of figure \ref{Fig3}). All the following measurements are performed in the saturation plateau (see vertical lines in figure \ref{Fig3}) where we have a large population in the spin states $\pm$1 corresponding to $k_BT_{eff}^{Cr}\gg D_0$. In this hight excitation regime no shift of the exciton line is detected excluding thermalization of the phonons at the excitation spot and an increase of the lattice temperature T$_0$.

\section{Spin relaxation in the presence of non-equilibrium phonons.}

As shown in the previous section, under high energy excitation non-equilibrium phonons controls the Cr spin temperature. This permits to populate the spin states S$_z=\pm1$. However, non-equilibrium phonons can remain in the sample hold on a cold finger in vacuum, a few $\mu$s after the end of the non-resonant pulse \cite{Msall1997}. Under hight intensity phonons could even be created by the resonant excitation used for the pumping (absorption in other QDs within the laser focal spot).

\begin{figure}[hbt]
\centering
\includegraphics[width=3.5in]{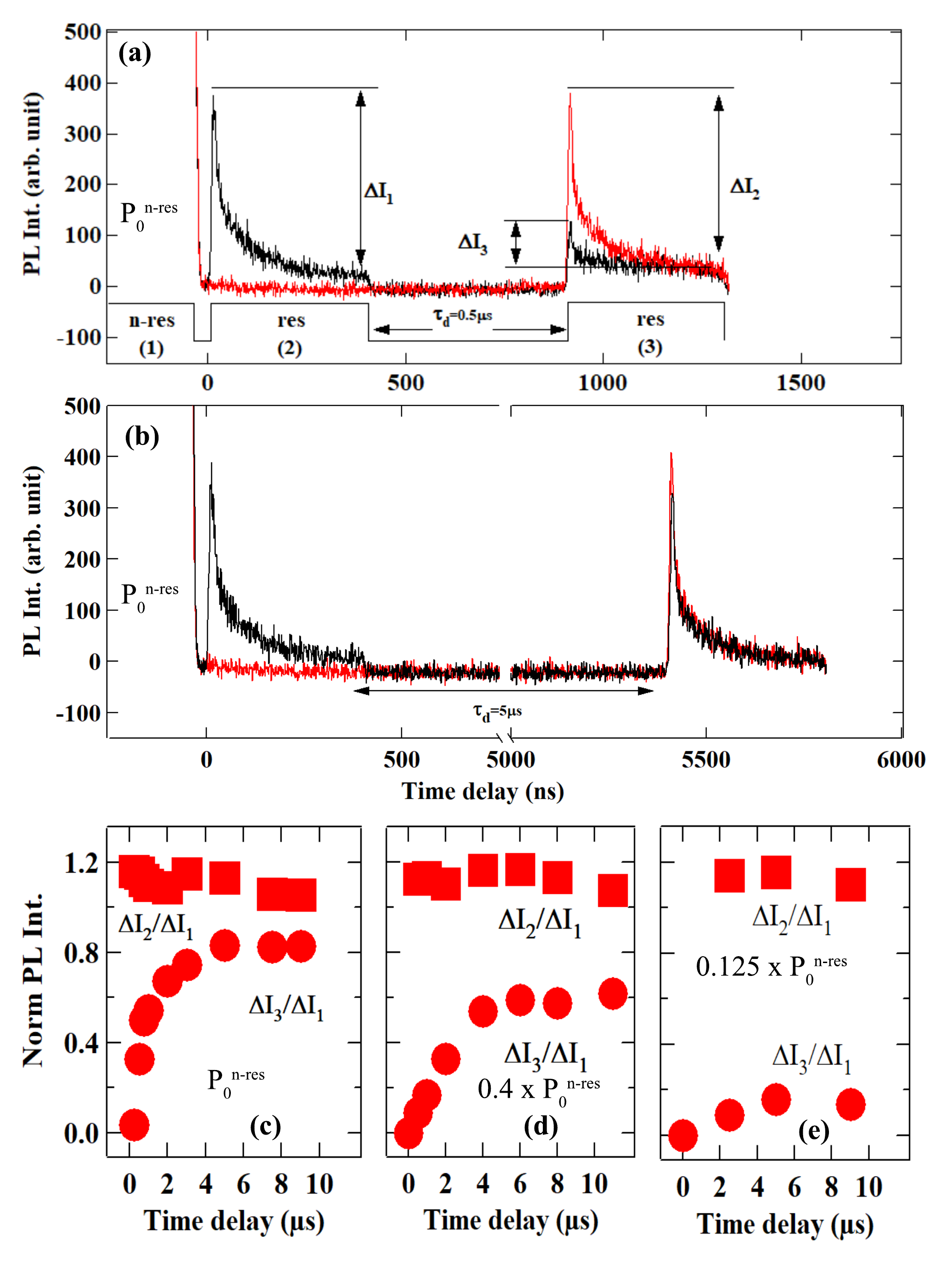}
\caption{Non-resonant excitation power dependence of the Cr spin relaxation time measured in a three pulses resonant pumping experiment: pulse (1) populates the spin sates $\pm1$, pulse (2) initializes the Cr spin (empties +1 or -1) and pulse (3) probes the relaxation in the dark. Pulse (2) can be suppressed to probe the cooling of the Cr using only pulse (3). (a) and (b): Pumping transients obtained at hight non-resonant excitation power (P$_0^{n-res}$) for two different dark times $\tau_d$. (c), (d) and (e), evolution as a function of $\tau_d$ of the normalized amplitudes $\Delta I_2/\Delta I_1$ and $\Delta I_3/\Delta I_1$ for three different non-resonant excitation intensities P$_0^{n-res}$, 0.4 x P$_0^{n-res}$ and 0.125 x P$_0^{n-res}$ respectively.}
\label{Fig4}
\end{figure}

To analyse the influence of non-equilibrium phonons on the spin dynamics of an isolated Cr atom we developed a three pulses optical pumping experiment. A first non-resonant pulse (pulse (1) in figure \ref{Fig4}) is used to increase the effective temperature of the Cr spin and populate the high energy spin states S$_z$=$\pm$1. The amount of non-equilibrium phonons generated in the sample during this non-resonant excitation is controlled by the excitation power P$^{n-res}$. Then, the resonant circularly polarized pulse 2 is tuned to the high energy line of X-Cr to initializes the Cr spin ({\it i.e.} empty the spin state S$_z$=+1 for a $\sigma$+ excitation or S$_z$=-1 for a $\sigma$- excitation).

To probe the Cr spin relaxation after the spin pumping sequence, a third resonant pulse (pulse 3) with the same energy and polarization as pulse 2 is sent after a variable dark time $\tau_d$ (figure \ref{Fig4}(a) and (b)). The amplitude of the transient during this pulse, $\Delta$I$_3$, depends on how the high energy Cr spin state has been populated during the dark time. Alternatively, pulse 2 can be suppressed in the excitation sequence and the amplitude of the pumping signal obtained during  pulse 3 used to monitor the cooling of the Cr spin during the dark time. For each measurements, the pumping transients are recorded on the resonant PL arising from the central line C.

The main results of these time resolved optical pumping experiments are summarized on figure \ref{Fig4}. The dependence on $\tau_d$ of the amplitude of the transients $\Delta$I$_2$ and $\Delta$I$_3$ (normalized by $\Delta$I$_1$) are presented on figure \ref{Fig4}(c), (d) and (e) for three different values of the intensity of the non-resonant initial heating pulse 1.

First we can notice that after an initial pumping (pulse 2), the effective temperature of the Cr spin reached after 10 $\mu$s (the time range studied here) proportional to $\Delta$I$_3$, strongly depends on the intensity of the initial non-resonant pulse P$^{n-res}$. Whereas for a high excitation power the high effective temperature of the Cr spin is almost completely restored after a 10 $\mu$s dark time, at low power the amplitude of the second transient $\Delta$I$_3$ remains weak even at long delay.

The speed of the Cr spin relaxation after the initial optical pumping ({\it i.e.} spin heating) also depends on the presence of acoustic phonons generated during the non-resonant excitation (initial heating pulse 1). At high P$^{n-res}$, the amplitude of the pumping transient is restored after a few hundreds $ns$. At low non-resonant excitation power, the spin relaxation can be hardly observed in the studied time range.

If the first resonant pumping pulse is suppressed we observe that, whatever the considered non-resonant pump power P$^{n-res}$, no significant cooling of the spin occurs during the time scale considered here ($\Delta I_2/\Delta I_1$ remains close to 1 \cite{zizi}). Consequently, the Cr spin population reached after 10 $\mu$s can be very different with and without the presence of the first pumping pulse 2. This is especially the case at low non-resonant excitation power: In the absence of initial optical pumping, a high spin temperature is conserved during the dark time whereas if a pumping is initially performed, the population of the spin state which was emptied remains very weak.

\begin{figure}[hbt]
\centering
\includegraphics[width=3.25in]{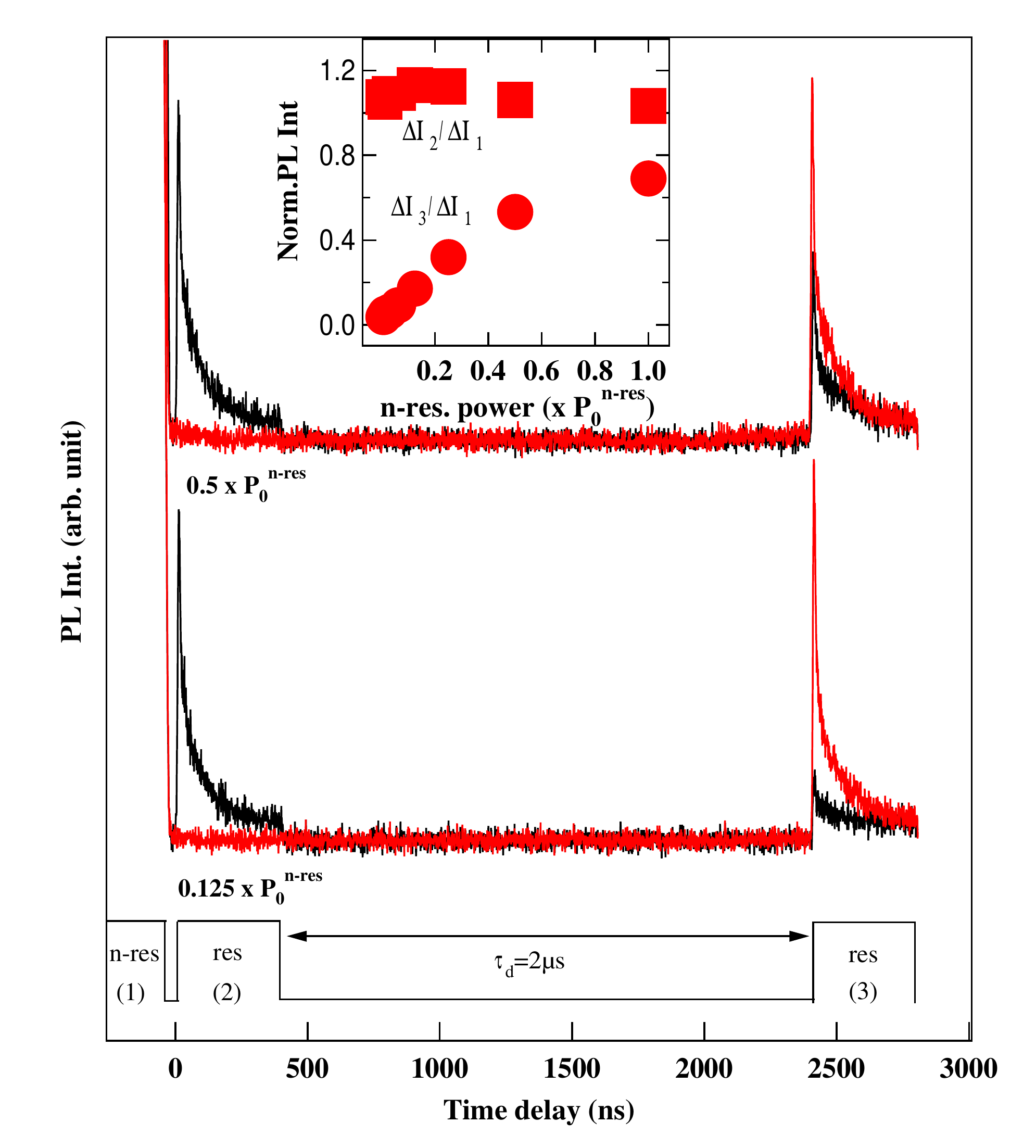}
\caption{Evolution of the resonant pumping transients as a function of the power of the non-resonant pulse 1, $P^{n-res}$, for a fixed dark time $\tau_d$=2$\mu$s. The inset shows the $P^{n-res}$ dependence of the normalized amplitudes $\Delta I_2/\Delta I_1$ and $\Delta I_3/\Delta I_1$.}
\label{Fig5}
\end{figure}

The influence of the non-equilibrium phonons on the Cr spin relaxation can also be revealed by changing the intensity of the non-resonant heating P$^{n-res}$ for a fixed dark time $\tau_d$. The evolution of the normalized amplitude of the pumping transients $\Delta I_2/\Delta I_1$ and $\Delta I_3/\Delta I_1$ are presented in figure \ref{Fig5} for $\tau_d$=2$\mu$s and a variable power of pulse 1. When only the heating pulse is used (pulse 2 off) the normalized transient $\Delta I_2/\Delta I_1$ remains close to 1 \cite{zizi} over the full power range used here: the Cr spin population of the spin states S$_z=\pm$1 is conserved during the dark time independently of the amount of generated non-equilibrium phonons.

If the initial pumping is performed (pulse 2 on), the pumping transient $\Delta I_3$ is not observed after the dark time at low P$^{n-res}$. By increasing P$^{n-res}$, the normalized amplitude of the second transient $\Delta I_3/\Delta I_1$ increases and at high excitation the initial effective temperature of the Cr is almost completely restored after the 2 $\mu$s dark time. This power dependence at fixed delay shows that with the decrease of the density of non-equilibrium phonons, $\Delta I_3$ decreases more quickly than $\Delta I_1$ whereas the ratio $\Delta I_2/\Delta I_1$ remains almost unchanged.

$\Delta I_1$, proportional to the population of the spin states S$_z$=+1 or S$_z$=-1 at the end of the heating pulse 1, is controlled by the steady state number of non-equilibrium phonons generated at the QD location during the non-resonant pulse. For a fixed $\tau_d$, $\Delta I_3$ is controlled by the spin-lattice coupling which can depend on the number of non-equilibrium phonons remaining during the dark time after the end of the non-resonant pulse. Their influence is particularly important in the first few $\mu$s where the population of the spin states S$_z$=+1 or S$_z$=-1 increases. Then the S$_z$=+1 or S$_z$=-1 spin population stabilizes at longer delay where most of the non-equilibrium phonons have been evacuated from the sample.

This ensemble of measurements shows that the speed and the efficiency of the Cr spin heating in the dark strongly depends on the presence of the non-equilibrium phonons optically generated during an initial spin heating pulse. One can however find low intensity excitation conditions for which a large population of the spin states $S_z=\pm$1 is obtained after the heating pulse and no significant relaxation is observed during a dark time of a few $\mu$s (Figure \ref{Fig5}, bottom curves).

\section{Discussion on the mechanism of Cr spin relaxation.}

We now discuss the mechanism that explains the observed Cr spin dynamics. In the absence of free carriers, the spin relaxation for an isolated atom with large magnetic anisotropy in a strained semiconductor host is controlled by the interaction with the lattice. In the case of a Cr atom which carries an orbital momentum, the spin-lattice coupling is induced by a modulation of the crystal field by the strain field of acoustic phonons. This modulation of the local electric field acts on the orbital part of the wave function of the magnetic atom and, combined with the spin-orbit coupling, can produce transitions between spin levels \cite{Abragam}.

Modulation of the crystal field by phonons can give rise to a direct spin relaxation process in which a single phonon is emitted or absorbed during the spin transition. This process is dominant at low temperature where phonons can only be present at the energy of the transitions between the two low energy levels, the Cr spin states S$_z$=0 and S$_z$=$\pm$1 separated by an energy D$_0$. The efficiency of this process depends on the temperature and a T$_{0}$ dependence is expected for the relaxation rate for $k_BT_{0} \gg D_0$.

At higher temperature when higher energy phonon modes are populated or alternatively in the presence of high energy non-equilibrium phonons, it is also possible for a Cr atom to absorb and emit two high energy phonons to perform the transition between S$_z$=0 and S$_z$=$\pm$1 \cite{Scott1962}. These are indirect processes that can easily be faster than the direct transition because of the larger density of states for phonons at high energy \cite{Abragam,Scott1962}.

Different kinds of indirect processes are possible \cite{Abragam,Scott1962}. If an intermediate high energy state is present at an energy comparable with the available acoustic phonons two processes can occur: the two-phonons Orbach process associated with a real absorption followed by an emission of phonons by the intermediate state and a two-phonons Raman process with virtual phonon transitions. The probability of these two processes depend on the density of phonons at the energy of the excited state. Both can occur simultaneously but when the energy of the intermediate state is larger than the energy of the continuum of phonons, only the Raman process is possible. In the absence of high energy intermediate state, a Raman transition with a virtual absorption and emission of high energy phonons is also possible directly between the two ground states.

All these processes can occur simultaneously but for an energy of the intermediate state larger than the energy of the continuum of phonons or in the absence of high energy intermediate state, only the Raman process is possible. These two-phonon processes strongly depend on the temperature T$_{0}$ (a T$_{0}^7$  dependence is for instance predicted for the transition rate associated with a two-phonon Raman process \cite{Abragam}) or on the density of high energy non-equilibrium phonons N$_{\omega}$ that are generated during a high energy optical excitation in a semiconductor.

In the presence of high energy phonons, these processes coexist and lead to a fast monotonic decrease of the spin-lattice relaxation time with the increase of the number of phonons ({\it i.e.} increase of lattice temperature or injection of non-equilibrium phonons) \cite{Abragam}. At low temperature and in the absence of excitation the relaxation is ultimately controlled by the single phonon direct process. A time dependence of N$_{\omega}$ as expected during the relaxation of non-equilibrium phonons, will result in a time varying spin-lattice coupling. Lets analyse how these processes and their time dependence under modulated optical excitation can explain the observed relaxation of the Cr spin.

During the high energy non-resonant pulse 1, a high density of non-equilibrium phonons is generated within the laser spot located above the QD \cite{Lafuente2018}. The spin-lattice relaxation time becomes very short and the Cr spin is heated by both direct and indirect phonons processes. Phonons generated within the laser spot travel in the sample but are not evacuated efficiently as the sample is hold under vacuum on a cold finger. Optically generated phonons remain in the sample at least a few $\mu$s \cite{Abragam, Msall1997}, scatter at the surface or on defects with a shift of their spectral distribution to lower energies progressively approaching the thermal bath.

At low non-resonant excitation power (0.125$\times P_0^{n-res}$ in figure \ref{Fig4}(e)), the non-equilibrium phonons are evacuated from the lasers spot and dispersed in the whole volume of the sample. Their density at the QD location is already very weak a few tens of $ns$ after the end of pulse 1. Just after the pulse, the fast two phonons process involving high energy phonons vanishes and only the one phonon process associated with low energy acoustic phonons remains. The spin lattice relaxation time gets closer to $\tau_{sl}(T_0)$ and do not lead to significant spin relaxation in the studied time range (less than 10 $\mu$s). As observed in the experiments, the population of the spin state which has been emptied by resonant optical pumping remains weak even at long delay (evolution of $\Delta I_3/\Delta I_1$). In the absence of initial optical pumping no significant cooling of the spin is observed during the first 10 $\mu$s (evolution of $\Delta I_2/\Delta I_1$) showing that $\tau_{sl}(5K)$ is larger than a few tens of $\mu$s.

The situation is completely different at high non-resonant excitation intensity ($P_0^{n-res}$ in figure \ref{Fig4}(c)). At high power, a high density of non-equilibrium phonons remains within the sample a few $\mu$s after the end of pulse 1. $\tau_{sl}$ remains shorter than the average non-equilibrium phonon lifetime during a few $\mu$s and lead to a significant spin heating after the end of the optical pumping by pulse 2. The Cr spin returns in a few hundreds $ns$ - one $\mu$s to a high effective temperature close to the spin temperature reached just after the non-resonant pulse 1.

In the intermediate non-resonant power range (0.4$\times P_0^{n-res}$ in figure \ref{Fig4}(d)), the density of non-equilibrium phonons after the pump pulse is weaker and the Cr spin returns, after relaxation, to an intermediate effective spin temperature lower than the initial temperature reached just after pulse 1.

A surprising behaviour is the absence of significant cooling of the Cr spin after pulse 1. This can be understood if the spin-lattice interaction time, which quickly decreases with the increase of the number of phonons, becomes shorter than the non-equilibrium phonon lifetime (a few $\mu$s) only when $k_BT_{eff}^{ph}\gg D_0$. With $\tau_{sl}(D_0/k_B)$ already larger than a few $\mu s$, $\tau_{sl}(T_{eff}^{ph})$ becomes shorter than the non-equilibrium phonon lifetime only at high effective temperature $T_{eff}^{ph}$. In this regime with $T_{eff}^{ph} \gg D_0/k_B$ the interaction with non-equilibrium phonons heats up the Cr spin after pumping but cannot lead to any significant cooling after the end of pulse 1. 

At long time delay, longer than what can be studied with the resonant pumping technique presented here, the Cr spin will return to the temperature of the lattice T$_0$ with a weak population of the states S$_z$=$\pm1$ and most of the population will relax on the spin ground state S$_z$=0. The time-scale of this relaxation is ultimately controlled by the direct single phonon process dominant at T$0$=5K and could take place in a few tens of $\mu$s ($\tau_{sl}(5K)$).

\section{Conclusion}

We demonstrated that the spin relaxation of an isolated Cr spin in a CdTe/ZnTe QD is strongly sensitive to the presence of non-equilibrium phonons. As a consequence, the Cr spin temperature and relaxation are both dependent on the optical excitation conditions used for the measurements. A spin-lattice coupling time decreasing with the increase of the density of high energy non-equilibrium phonons can explain the observed spin dynamics. One should therefore be particularly careful with the excitation conditions to be used for an efficient optical probing and control of a single Cr spin embedded in a QD.

We have however shown that the S$_z$=$\pm$1 spin states of the Cr can be significantly populated by a non-resonant excitation pulse and that efficient resonant spin pumping could be achieved to select and read the states within the S$_z$=$\pm1$ spin sub-space. We found excitation conditions where the prepared spin state can be preserved in the dark on a $\mu$s time scale. This opens the possibility to coherently control the $\lbrace-1;+1 \rbrace$ Cr spin $qubit$ with a resonant driving field in the GHz range. For example, one could use surface acoustic waves, which are particular non-equilibrium phonons travelling at the surface of a sample at a well controlled frequency \cite{Whiteley2019}, for a coherent mechanical driving of the $\lbrace-1;+1 \rbrace$ Cr spin $qubit$.

\begin{acknowledgements}{}

This work was realized in the framework of the Commissariat \`{a} l'Energie Atomique et aux Energies Alternatives (Institut Nanosciences et
Cryog\'{e}nie) / Centre National de la Recherche Scientifique (Institut N\'{e}el) joint research team NanoPhysique et Semi-Conducteurs. The work was
supported by the French ANR project MechaSpin (ANR-17-CE24-0024) and CNRS PICS contract No 7463. V.T. acknowledges support from EU Marie Curie grant
No 754303. The work in Tsukuba has been supported by the Grants-in-Aid for Scientific Research on Innovative Areas "Science of Hybrid Quantum
Systems" and for challenging Exploratory Research.

\end{acknowledgements}


\begin{thebibliography}{}
\bibitem{zizi} The tiny difference between $\Delta I_1$ and $\Delta I_2$ arises from a difference in the effective spin temperature of the exciton-Cr system during pulse 2 and pulse 3. A larger effective temperature during pulse 2 (just after the hight intensity non-resonant excitation) results in a slightly weaker transfer towards the low energy exciton-Cr state S$_z$=0 and weaker resonant PL. The amplitude of this effect is close to the experimental uncertainty. This effect does not have any consequence on the conclusion of the paper concerning the dynamics of the Cr spin alone in the absence of carriers.
\end{thebibliography}
\end{document}